\documentclass[reprint,prb,twocolumn,showpacs,superscriptaddress,aps,longbibliography]{revtex4-1}
\usepackage[utf8]{inputenc}
\usepackage{graphicx}
\usepackage[colorlinks=true,citecolor=blue]{hyperref}
\usepackage{amsmath}
\usepackage{amssymb}
\usepackage{soul}
\usepackage{gensymb}
\usepackage{amsfonts,amssymb,amsbsy}
\usepackage{siunitx}
\usepackage{float}
\usepackage{physics}
\usepackage{booktabs}
\usepackage[percent]{overpic}

\newcommand{\dIdV}{d$I$/d$V$~}
\usepackage{xcolor}
\usepackage{natbib}

\usepackage{lineno}

\begin{document}

\title{Moir\'e modulated quantum spin liquid candidate 1T-TaSe$_2$}

\author{Ziying Wang}
\email{Corresponding authors. Email: ziying.wang@aalto.fi, jose.lado@aalto.fi, robert.drost@aalto.fi}
\affiliation{Aalto University, Department of Applied Physics, 00076 Aalto, Finland}

\author{Adolfo O. Fumega}%
\affiliation{Aalto University, Department of Applied Physics, 00076 Aalto, Finland}

\author{Ana Vera Montoto}%
\affiliation{Aalto University, Department of Applied Physics, 00076 Aalto, Finland}

\author{Mohammad Amini}
\affiliation{Aalto University, Department of Applied Physics, 00076 Aalto, Finland}

\author{Büşra Gamze Arslan}
\affiliation{Aalto University, Department of Applied Physics, 00076 Aalto, Finland}

\author{Ale\v{s} Cahl\'ik}
\affiliation{Aalto University, Department of Applied Physics, 00076 Aalto, Finland}

\author{Yuxiao Ding}
\affiliation{Aalto University, Department of Applied Physics, 00076 Aalto, Finland}

\author{Jose L. Lado}%
\email{Corresponding authors. Email: ziying.wang@aalto.fi, jose.lado@aalto.fi, robert.drost@aalto.fi}
\affiliation{Aalto University, Department of Applied Physics, 00076 Aalto, Finland}

\author{Robert Drost}
\email{Corresponding authors. Email: ziying.wang@aalto.fi, jose.lado@aalto.fi, robert.drost@aalto.fi}
\affiliation{Aalto University, Department of Applied Physics, 00076 Aalto, Finland}

\author{Peter Liljeroth}
\affiliation{Aalto University, Department of Applied Physics, 00076 Aalto, Finland}

\begin{abstract}
Quantum spin liquids are quantum phases of matter featuring collectively entangled states and emergent fractional many-body excitations. While methods exist to probe three-dimensional quantum spin liquids experimentally, these techniques lack the sensitivity to probe two-dimensional quantum spin liquids. This seriously hampers the study of potential monolayer quantum spin liquid candidates such as $\alpha$-RuCl$_3$ and 1T-TaSe$_2$. Scanning tunneling microscopy (STM) and spectroscopy (STS) have recently been suggested as promising probes of the quantum spin liquid state, as they can access the spinon spectrum through inelastic tunneling spectroscopy (IETS). In this work, we employ this approach on the quantum spin liquid candidate material 1T-TaSe$_2$ and directly measure its low-energy inelastic excitations.
We observe the emergence of a $\sqrt{3}\times\sqrt{3}$ reconstruction driven by the substrate, equivalent spectroscopy across all spin sites 
and coexistence of zero and finite energy excitations. We show that these observations are consistent with a modulated $\sqrt{3}\times\sqrt{3}$ spin liquid ground state. Our results demonstrate that IETS provides a powerful route to obtain atomic-scale insight into the magnetic excitations of two-dimensional materials, allowing to explore the effects of moir\'e modulations on potential quantum liquid phases.

\end{abstract}

\date{\today}

\maketitle 

Quantum spin liquids (QSLs) are exotic states of matter that arise
in interacting quantum spin systems that do not order even at very low temperatures. They are predicted to exhibit exotic phenomena including fractionalized excitations and topological order. \cite{savary2016quantum,law20171t,RevModPhys.89.025003,broholm2020quantum} Beyond their fundamental appeal, QSLs have been proposed for applications in fault-tolerant quantum computation\cite{PhysRevX.10.031014}, and doped QSLs may provide a route to high temprature unconventional superconductivity\cite{PhysRevX.6.041007}. However, the characterization of QSLs remains challenging even in the bulk, and many open questions remain.\cite{ broholm2020quantum,savary2016quantum,takagi2019concept,wen2019experimental} The magnitude of the measured signal with the most used experimental probes, inelastic neutron scattering, NMR, µSR, and thermal Hall transport \cite{shen2016evidence,klanjvsek2017high,kratochvilova2017low},
depends on the sample volume and the sensitivity is insufficient for two-dimensional (2D) systems. Consequently, studying  QSL candidates in two dimensions remains an open challenge.

Two-dimensional QSL candidate materials arise in magnetically frustrated systems, with frustrated interactions in $\alpha$-RuCl$_3$ and geometric frustration in 1T-TaS$_2$ being prominent examples. In bulk $\alpha$-RuCl$_3$, inelastic neutron scattering has revealed a continuum of spin excitations, a signature of a proximal QSL phase,\cite{banerjee2017neutron} featuring potential chiral modes.\cite{kasahara2018majorana} Theoretical work has indicated that monolayer $\alpha$-RuCl$_3$ might be even closer to an intrinsic Kitaev-type QSL than the bulk material.\cite{PhysRevLett.123.237201,RuCl3_valenti_2025}
In 1T-TaS$_2$, charge density wave reconstruction results in a 13 Ta atom unit cell that hosts an unpaired spin.\cite{law20171t,Vano2021} Nuclear magnetic resonance, neutron scattering and muon spin resonance experiments have shown zero static magnetic moments down to 20 mK \cite{klanjvsek2017high,kratochvilova2017low,ribak2017gapless} with a residual linear term in thermal conductivity, which implies the possible presence of a QSL state\cite{yu2017heat,murayama2020effect}. In a closely related 2D system, monolayer 1T-TaSe$_2$, STM-based quasiparticle interference data and Kondo effect provide evidence of the potential presence of a spinon Fermi surface with gapless spinon excitations.\cite{ruan2021evidence,chen2022} 
2D QSL candidates should benefit from their ultimate tunability via gating, moir\'e patterns, and strain.\cite{wang2015physical}  The presence of proximal QSL phases in these systems highlights the need for new tools specialised in probing 2D materials.

Inelastic electron tunnelling spectroscopy (IETS) provides an experimental
strategy to probe two-dimensional quantum spin  liquids \cite{PhysRevLett.125.267206,PhysRevResearch.2.033466,PhysRevB.107.054432,PhysRevB.109.035127},
analogously to magnon excitations in a ferromagnet\cite{Ghazaryan2018,Ganguli2023} or spin-flip excitations in isolated spins \cite{Heinrich2004,Ternes2015}. In this work, we probe the spin liquid candidate monolayer 1T-TaSe$_2$ by IETS with a scanning tunneling microscope (STM). The low-energy IETS spectra show symmetric features arising from many-body excitations inside the Mott gap of monolayer 1T-TaSe$_2$. These excitations have a $\sqrt{3}\times\sqrt{3}$ modulation in real space and they stem from a substrate-induced moir\'e reconstruction of the many-body ground state in 1T-TaSe$_2$, featuring zero energy and finite energy excitations.  Using many-body methods and auxiliary fermions, we show that the observed spectroscopy is consistent with a moir\'e-modulated QSL state. Our work shows that IETS is a powerful tool for probing many-body excitations in 2D materials at the atomic scale and points towards possibilities of moir\'e engineering of QSL states.

\begin{figure*}[ht!]
    \centering
    \includegraphics{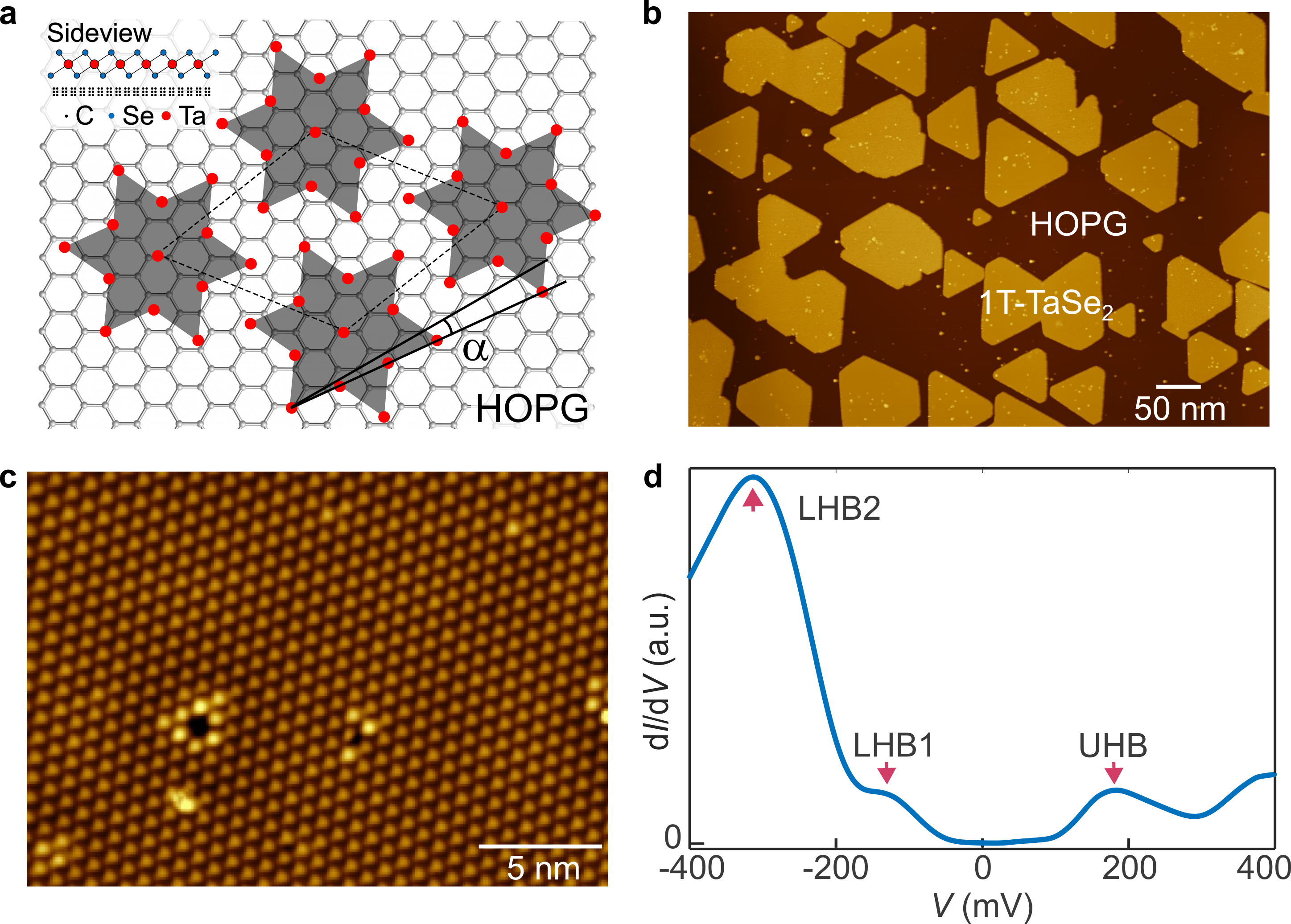}
    \caption{CDW and Mott insulator state in monolayer 1T-TaSe$_2$ grown by MBE and characterized by STM at 5 K. \textbf{a,} Schematic of monolayer 1T-TaSe$_2$ crystal and the star-of-David structure. The red, blue and black atoms represent Ta, Se and carbon atoms. The angle between HOPG and 1T-TaSe$_2$ determines the moir\'e structure.  \textbf{b,} A large-scale STM image of 1T-TaSe$_2$ monolayer islands on HOPG substrate. $V=1$ V, $I=20$ pA. \textbf{c,} A high-resolution STM topography image of the CDW superlattice. $V=-0.5$ V, $I=50$ pA. \textbf{d,} Representative tunnelling spectrum on 1T-TaSe$_2$. Arrows indicate the two lower Hubbard bands (LHB1 and LHB2) and the upper Hubbard band (UHB).}
    \label{Figure1}  
\end{figure*}

\subsection*{Moir\'e modulation in 1T-TaSe$_2$}

Monolayer 1T-TaSe$_2$ films were grown on freshly cleaved highly oriented pyrolytic graphite (HOPG) substrates via molecular beam epitaxy (MBE) (details in the methods section).\cite{chen2020strong,nakata2021robust} 1T-TaSe$_2$ layer contains one Ta atomic layer sandwiched between a pair of Se atomic layers, with each Ta atom coordinated by six Se atoms (Figure \ref{Figure1}a). The Se cage forms an octahedron in the 1T phase. Figure \ref{Figure1}b shows a large-scale topography of the MBE-grown TaSe$_2$ islands, which are predominantly 1T phase with high crystallinity and ultra-clean surfaces. The islands are nearly aligned with the HOPG substrate; the relative orientation between the atomic lattices of TaSe$_2$ and HOPG varies between $1^\circ$ and $3^\circ$ (see SI Figure 2). Figure \ref{Figure1}c shows a zoomed-in STM image, where the charge density wave (CDW) modulation is visible as a triangular lattice. The Fourier transformed (FT) image (SI Figure 1) confirms the CDW structure of $\sqrt{13} \times \sqrt{13}$ times atomic lattice with an angle of $13.9^\circ$ between the atomic and CDW lattices.\cite{chen2020strong,nakata2021robust} Tunneling spectroscopy in Figure \ref{Figure1}d shows the lower (-130 mV) and the upper Hubbard bands (180 mV) separated by the Mott gap, consistent with previous works.\cite{chen2020strong,nakata2021robust} The orbital textures at different energies can be resolved via \dIdV maps (SI Figure 3), which are consistent with previous findings.\cite{chen2020strong} 

\begin{figure*}[ht!]
    \centering
    \includegraphics[width =.9 \textwidth]{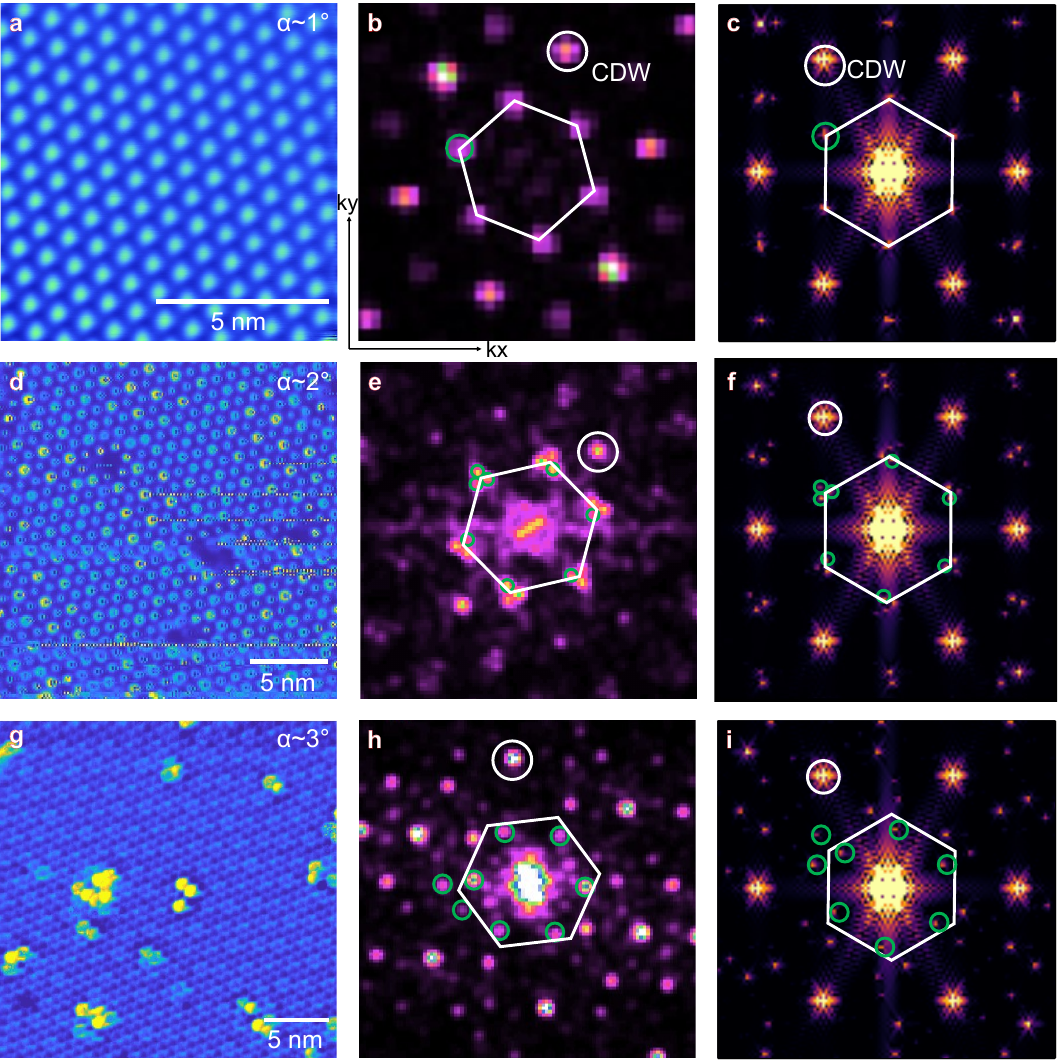}
    \caption{Moir\'e modulation in 1T-TaSe$_2$ on HOPG substrate at different lattice orientations. \textbf{a,d,g,} \dIdV maps (extracted from grid data, at biases around LHB1) on different 1T-TaSe$_2$ islands with atomic lattices rotated by $1^\circ$, $2^\circ$, and $3^\circ$ relative to the HOPG lattice. Besides CDW superlattices, new domain-like modulations are observed on $2^\circ $ and $3^\circ $ islands. \textbf{b,e,h,} FT images of the corresponding \dIdV maps. The white circles highlight one of the CDW peaks and the hexagons highlight the CDW Brillouin zone. At or near the K and K' points of the CDW Brillouin zone, new sets of peaks highlighted by green circles correspond to the moir\'e pattern. \textbf{c,f,i,} Simulated FT images of moir\'e with relative angles around $1^\circ $, $2^\circ$, and $3^\circ$ from HOPG.}
    \label{Figure2}
\end{figure*}

The different 1T-TaSe$_2$ monolayer islands have slightly different orientations with respect to the HOPG substrate, and this results in different moir\'e periodicities on the sample. When the angle between the two atomic lattices is $1^\circ $, the moir\'e pattern forms a  $\sqrt{3} \times \sqrt{3}$ reconstruction with respect to the CDW (later written as $\sqrt{3} $ for simplicity), as shown in Figure \ref{Figure2}a-b. Fourier transforms (FTs) of the topographies clearly show the peaks corresponding to the moir\'e pattern at the K and K' points of the CDW Brillouin zone (see Fig.~\ref{Figure2}b). This result agrees well with the simulation when the lattice angle difference between the 1T-TaSe$_2$ and HOPG is at $1^\circ $ (Figure \ref{Figure2}c). As the lattices rotate with respect to each other, the moir\'e shifts from the originally commensurate  $\sqrt{3}$ modulation to an incommensurate superlattice, manifesting a topographically domain-like structure (Figure \ref{Figure2}d). The FT peaks of the moir\'e (Figure \ref{Figure2}e) also rotate slightly from the $K$ and $K'$ points of the CDW Brillouin zone, forming a trimer-like structure with the moir\'e peaks from the second CDW Brillouin zone. The simulation with $2^\circ$ lattice rotation shows comparable features (Figure \ref{Figure2}f) to the experiments. When the lattice rotation mismatch increases to $ 3^\circ$, the moir\'e modulation becomes more obvious in the STM images (Figure \ref{Figure2}g) and the corresponding FT peaks are further separated compared to the rotation $2^\circ$ (Figure \ref{Figure2}h). The simulation at this angle in Figure \ref{Figure2}i also matches the experimental results. More systematic simulation results of the angle-dependent moir\'e evolution of the topographies and the corresponding FTs are shown in SI Figure 4. Locally, even when the rotational alignment is broken, the modulation periodicity is still very close to $\sqrt{3}$. However, the phase of this modulation w.r.t.~the CDW lattice shifts over longer length scales. 

\begin{figure}[ht!]
    \centering
    \includegraphics[width = \columnwidth]{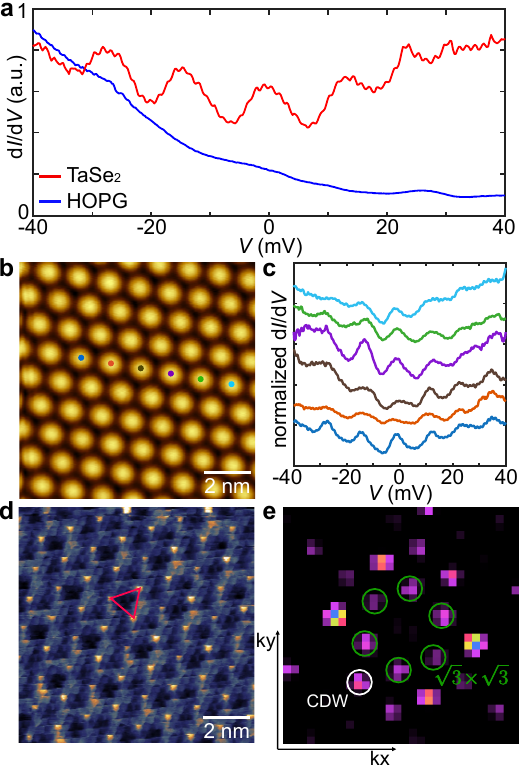}
    \caption{Low energy inelastic excitations in 1T-TaSe$_2$. \textbf{a,} In-gap inelastic excitations of monolayer 1T-TaSe$_2$ (red line). Spectrum obtained with the same tip on HOPG as a comparison (blue line). \textbf{b,} STM image over serveral $\sqrt{3} $ moir\'e cells.  $V= -0.5$ V, $I=100$ pA.  \textbf{c,} \dIdV at neighbouring CDW centers indicated in panel b. \textbf{d,} A detailed constant-height \dIdV map of $\sqrt{3} $ moir\'e modulation. $V= -15$ mV, $I=400$ pA. Red triangles highlight the moir\'e trimerization. \textbf{e,} FT of the data in panel d showing the $\sqrt{3} \times \sqrt{3}$ reconstruction.}
    \label{Figure3}
\end{figure}

\subsection*{Low energy excitation spectra of 1T-TaSe$_2$}

We probed many-body excitations in monolayer 1T-TaSe$_2$ by low-energy IETS measurements at 5 K. Inelastic tunnelling on magnetic systems causes spin-flip processes ($\Delta S = 1$), and these additional tunneling channels contribute to the measured d$I$/d$V$, which can then be used to probe the fundamental magnetic excitations of the sample.\cite{kim2006distinct,mross2011charge,Ternes2015} The red curve in Figure \ref{Figure3}a is a typical IETS spectrum collected on monolayer 1T-TaSe$_2$ and it shows a zero-bias peak and two pairs of symmetric peaks 
associated with many-body excitations. The blue curve in Figure \ref{Figure3}a is acquired on the HOPG substrate with the same tip and it shows no bias-symmetric features as expected. 
We recorded several IETS spectra at 6 neighboring CDW centers, whose positions are indicated in Figure \ref{Figure3}b, and the corresponding spectra are shown in Figure \ref{Figure3}c (further data in SI Section 3A and Figure S7). The spectroscopy taken at the different CDW sites consistently shows 5 resonant peaks, including a zero-energy peak, across the different unit cells, so the moir\'e does not break a symmetry that would lead to inequivalent excitations in the different CDW sites. 

Figure \ref{Figure3}d shows a constant-height \dIdV map at a bias of $V=-15$ mV corresponding to one of the resonances in the \dIdV spectra (other low-bias constant-height DOS maps are shown in the SI Figure S8). The \dIdV intensity shows the spatial variation of the intensity of the low-bias spectral features, and it is essentially equal in the middle of the CDW sites. However, there is a clear modulation in between the sites that reflects the $\sqrt{3}$ modulation. This is better visualized in the FT of the map shown in Fig.~\ref{Figure3}e. This implies a $\sqrt{3} \times \sqrt{3}$ modulation of the interactions between the spins $1/2$ of the CDWs, and therefore will influence the magnetic ground state of the system. These experimental observations put constraints on the symmetry of the model that we will use to interpret the low energy excitations. The modulation corresponds to a trimerization between the sites (highlighted by the red triangle in Fig.~\ref{Figure3}d), where the 3 sites are equivalent. 

\begin{figure*}[ht!]
    \centering
    \includegraphics[width = \textwidth]{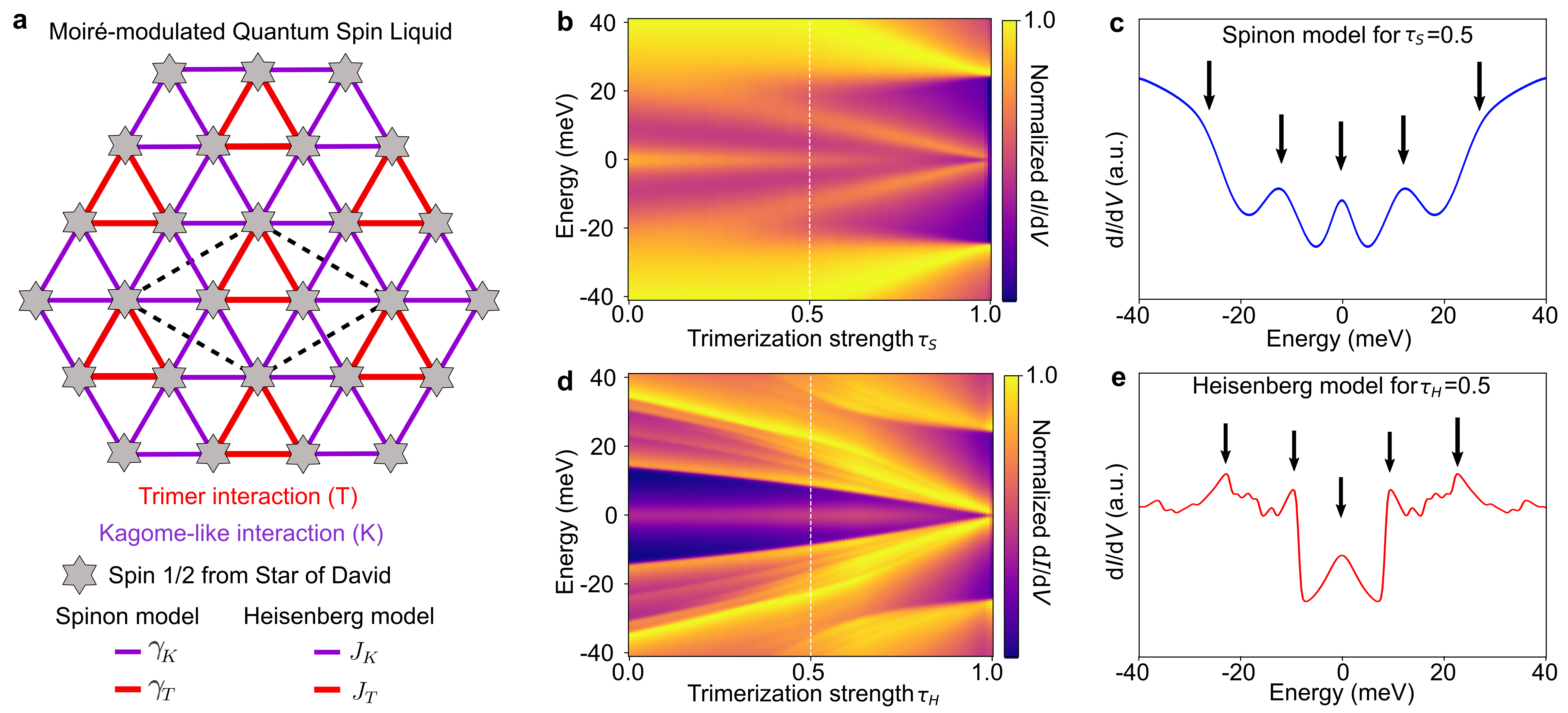}
    \caption{Theoretical modeling of a moir\'e-modulated QSL state in 1T-TaSe$_2$.  \textbf{a}, Sketch of the spinon and Heisenberg models considered for a moir\'e-modulated QSL. The grey stars represent localized spins 1/2. The moir\'e introduces an effective $\sqrt{3}\times\sqrt{3}$ modulation on the triangular lattice, creating a trimerization of the spins 1/2.
    Spinon and many-body models with two inequivalent coupling parameters in the TaSe$_2$ lattice. \textbf{b,d} Simulated dI/dV as a function of energy and trimerization strength values for the spinon and Heisenberg models. \textbf{c,e} Simulated dI/dV with $ \tau_S = \tau_H = 0.5$ for spinon model and many-body models, respectively.} 
    \label{Figure4}
\end{figure*}

\subsection*{Understanding the low energy excitations in 1T-TaSe$_2$}

We now proceed to interpret the low-energy excitations observed in our experiment.
1T-TaS$_2$ and 1T-TaSe$_2$ monolayers are potential two-dimensional QSL candidates\cite{law20171t,klanjvsek2017high,kratochvilova2017low,ribak2017gapless,yu2017heat,murayama2020effect,ruan2021evidence},
due to the intrinsic frustration of the triangular lattice formed 
by the CDW. Both 1T-TaSe$_2$ and 1T-TaS$_2$ monolayers are correlation-driven Mott insulators and show zero net magnetization down to low temperatures\cite{chen2020strong,wilson1975charge,tian2024dimensionality}. The low-energy in-gap excitations thus carry information about the magnetic ground state of  1T-TaSe$_2$.
In particular, inelastic excitations of a Mott state reflect $S=1$ spin excitations associated with
the spin structure factor $A(\omega)=\langle\Omega|S^-_n\delta (\omega-H+G_{\Omega}) S^+_n| \Omega\rangle$ of the ground state. 
Purely magnetic inelastic excitations (e.g.~spin-flip excitations) would result in a monotonically increasing \dIdV signal (as \dIdV$\propto$ integrated inelastic spectral function). 
This is clearly in contrast to our experiments, suggesting the presence of an additional channel, potentially a Kondo contribution\cite{Ternes2015}. 
This would indicate that the zero-bias feature arises from Kondo coupling between the magnetic moments hosted in the CDW unit cells and the underlying HOPG substrate. 
The finite-bias features then arise from the Kondo-assisted magnetic inelastic excitations of the 1T-TaSe$_2$ layer.\cite{PhysRevLett.96.176801,PhysRevLett.101.217203,Krane2025,PhysRevB.101.125405,kezilebieke2019observation,Sun2025}  
While we cannot directly rule out a magnetically ordered ground state solely on inelastic spectroscopy, 
the presence of a zero-bias Kondo feature suggests that an ordered state is not present (see SI Section S5 and Fig.~S11 for discussion on a moiré modulated magnet).
In contrast, QSL ground state can coexist with Kondo screening from the HOPG
metallic states\cite{PhysRevB.97.235117,Zhao2019}, a scenario compatible with our observations.
An alternative interpretation that could explain the zero-bias resonance would be a paramagnetic state, where each spin $1/2$ would behave as an independent impurity coupled to the metallic substrate. However, the higher-bias resonances would not be present
in this scenario (see SI Section S13 for a discussion on other alternatives).
All this suggests that the observed excitations may stem from a QSL state. 

Our experimental observations within the framework of a modulated QSL
are rationalized by a Hamiltonian for the CDW spin sites taking the form $H = \sum_{ij} J_{ij} \vec S_i \cdot \vec S_j$, where the antiferromagnetic exchange coupling $J_{ij}$ follows the $\sqrt{3}\times\sqrt{3}$ experimental modulation created by the moir\'e (Figure \ref{Figure3}d) and each $\vec S_n$ is localized in a 13-Ta star of the CDW. 
Figure \ref{Figure4}a shows a schematic of the Hamiltonian model for the modulated Heisenberg magnet. The moir\'e introduces $\sqrt{3}\times\sqrt{3}$ modulation on the first neighbor magnetic exchange interactions, $J_T$ couples the spins $1/2$ in trimers and $J_K$ provides a kagome-like interaction in the limit of $J_T=0$ (see Fig.~S9 for this limit in the SI). 
We define the strength for the trimerization as $\tau_H=1-J_K/J_T$ ($\tau_H=1$ in the limit of full trimerization).
In a QSL state, the spin Hamiltonian
can be solved with an auxiliary fermion Abrikosov replacement\cite{savary2016quantum}
$S^\alpha_n = \frac{1}{2}\sum_{s,s'}\sigma^\alpha_{s,s'} f^\dagger_{n,s} f_{n,s'}$ with the constraint $\sum_{s,s'} f^\dagger_{n,s} f_{n,s'} = \mathcal{I}$. Using the previous replacement on the spin Hamiltonian with a fixed point approximation leads to an effective spinon Hamiltonian of the form $H = \sum_{ij,s} \gamma_{ij} f^\dagger_{i,s} f_{j,s}$, with $\gamma_{ij} = \gamma_K,\gamma_T$, inheriting the $\sqrt{3}\times\sqrt{3}$ moir\'e modulation and the trimerization strength as $\tau_S=1-\gamma_K/\gamma_T$ (Figure \ref{Figure4}a). 
In the QSL state, the spin structure factor $A(\omega,n)$ reflects two-spinon excitations, and as a result, an effective self-convolution of the spinon spectral function\cite{PhysRevLett.125.267206}.
The simulated \dIdV
obtained is obtained from including the spin-flip assisted tunneling
together with a Kondo contribution (see SI Section S6 for details).
We shown in Fig.~\ref{Figure4}b the simulated \dIdV from the
spinon model as a function of the trimerization strength $\tau_S$
including a Kondo contribution. As a complementary benchmark,
we show in Fig.~\ref{Figure4}d the siulated \dIdV obtained
from the exact solution of the quantum Heisenberg model for
a finite lattice of 2D spin clusters as a function of the trimerization strength $\tau_H$.
As shown in Figs.~\ref{Figure4}b and \ref{Figure4}d, the ratio of trimerization has a strong effect on the peaks of
the simulated \dIdV. From Figures \ref{Figure4}c and \ref{Figure4}e, at 
strong moir\'e modulation ($ \tau $$_S$ and $ \tau $$_H \sim 0.5$), the 
simulated \dIdV for both models match quite well with the experimental
data (Figure \ref{Figure3}a). These findings thus suggest that the
experimental excitation spectrum in 1T-TaSe$_2$ is consistent
with a moir\'e-modulated QSL model.
In particular, the five resonant peaks are well captured by both
the spinon and quantum Heisenberg models, pointing out the crucial effect of the moir\'e modulation. 

\section*{Conclusions}

In this work, we have shown an emergence of a moir\'e induced modulation in the QSL candidate 1T-TaSe$_2$.
This moir\'e pattern plays an important role in modulating the low-energy inelastic excitations. 
Specifically, the inelastic modes feature a zero bias peak 
coexisting with higher energy excitations,
and equivalent spectroscopy on all sites
forming the $\sqrt{3}\times\sqrt{3}$
reconstructed structure.
We demonstrate that these experimental observations 
are consistent with a $\sqrt{3}\times\sqrt{3}$-modulated
QSL ground state in 1T-TaSe$_2$. 
Our work provides a strategy to explore QSL phases with STM, taking advantage of the signatures that moir\'e modulations can imprint on the excitations of these highly entangled quantum phases.

\section*{Methods}

\subsection*{Sample preparation} 

1T-TaSe$_2$ was grown by molecular beam epitaxy (MBE) on highly oriented pyrolytic graphite (HOPG) under ultra-high vacuum conditions (UHV, base pressure $\sim1\times10^{-10}$ mbar). HOPG crystal was cleaved and subsequently out-gassed at $\sim600^\circ$C in UHV. High-purity Ta was evaporated from an electron-beam evaporator. Se was evaporated from a Knudsen cell using Se powder ($99.9\%$, Merck). Before growth, the flux of Ta was calibrated on an Au(111) at $\sim1$ monolayer per hour. The sample was grown in a Se background pressure of $\sim1\times10^{-8}$ mbar and the growth duration was 30 minutes. Before the growth, the HOPG substrate temperature was stabilized at $\sim550^\circ$C. 

\subsection*{STM measurements}

After the preparation, the sample was inserted into the low-temperature STM (Createc LT-STM) connected to the same UHV system, and subsequent experiments were performed at $T = 5$ K. STM images were taken in the constant-current mode. d$I$/d$V$ spectra were recorded by standard lock-in detection while sweeping the sample bias in an open feedback loop configuration, with a peak-to-peak bias modulation specified for each measurement and at a frequency of 757 Hz. For IETS measurement, the modulation is 10mv.

\section*{Acknowledgements}
This research made use of the Aalto Nanomicroscopy Center (Aalto NMC) facilities and was supported by the Research Council of Finland Projects Nos.~371757, 369367, and 347266, EU Horizon Europe Marie Skłodowska-Curie Actions 101154353 and 101109672, ERC AdG GETREAL (no.~101142364), and ERC CoG ULTRATWISTROICS (no.~101170477). We acknowledge the financial support of the Finnish Ministry of Education and Culture through the Quantum Doctoral Education Pilot Program (QDOC VN/3137/2024-OKM-4), the Research Council of Finland through the Finnish Quantum Flagship project (358877, Aalto University) and the computational resources provided by the Aalto Science-IT project.

\section*{Contributions}
Z.W., R.D., and P.L. initiated and conceived this project. Z.W., M.A., B.G.A., R.D.,and A.C. carried out the STM/STS measurements. Z.W., B.G.A. and M.A. performed the sample growth. A.O.F., A.V.M., and J.L.L. performed the theoretical modelling. R.D. and Y.D. performed coding and data analysis. Z.W. wrote the manuscript with feedback from all the co-authors.

\bibliography{qsl}

\end{document}